\documentclass[lettersize,journal]{IEEEtran}
\usepackage{amsmath,amsfonts,amssymb}
\usepackage{algorithmic}
\usepackage{algorithm}
\usepackage{array}
\usepackage[caption=false,font=normalsize,labelfont=sf,textfont=sf]{subfig}
\usepackage{textcomp}
\usepackage{stfloats}
\usepackage{url}
\usepackage{verbatim}
\usepackage{graphicx,color}
\usepackage{cite}
\usepackage[normalem]{ulem}
\usepackage{xcolor}
\usepackage{epstopdf}
\newcommand{\f}[2]{\frac{#1}{#2}}
\newcommand{\pd}[2]{\f{\partial #1}{\partial #2}}
\newcommand{\pdd}[3]{\f{\partial^2 #1}{\partial #2 \partial #3}}
\begin{document}

%*** page limit: 10 pages!
%% *** update references
%% *** why do we need to solve G(p) = epsilon rather than G(p) = 0???
%%      -need better understanding/explanation
%%      -idea: default numerical precision is too small (i.e. requires too high
%%      accuracy and duration of numerical integration) so we need to increase
%%      the numerical tolerance (which we do artificially)

\title{Determining Disturbance Recovery Conditions by \\
  Inverse Sensitivity Minimization}

%% To specify the authors when (number of affiliations <= 2)
\author{
  \IEEEauthorblockN{Michael W. Fisher, {\it Member, IEEE} \hspace{1cm}
    Ian A. Hiskens, {\it Life Fellow, IEEE}}

\thanks{
  M. Fisher, michael.fisher@uwaterloo.ca, is with the Department of
  Electrical and Computer Engineering,
  University of Waterloo, 200 University Avenue West, Waterloo, ON N2L 3G1,
  Canada.
  
  I. Hiskens, hiskens@umich.edu, is with the Department of
  Electrical Engineering and Computer
  Science, University of Michigan, 1301 Beal Avenue, Ann Arbor, MI 48109, U.S.A.

  The authors gratefully acknowledge the
  contribution of the U.S.~National Science Foundation through grant
  ECCS-1810144.}

}

\maketitle

%*** note: our list of references is not very modern or long
%*** note: update references based on recent publications
% (add arxiv paper once available)

\begin{abstract}
Power systems naturally experience disturbances, some of which can damage equipment and disrupt consumers. It is important to quickly assess the likely consequences of credible disturbances and take preventive action, if necessary. However, assessing the impact of potential disturbances is challenging because many of the influential factors, such as loading patterns, controller settings and load dynamics, are not precisely known. To address this issue, the paper introduces the concept of parameter-space recovery regions. For each disturbance, the corresponding recovery region is the region of parameter space for which the system will recover to the desired operating point. The boundary of the recovery region establishes the separation between parameter values that result in trouble-free recovery and those that incur undesirable non-recovery. The safety margin for a given set of parameter values is defined as the smallest distance (in parameter space) between the given values and the recovery boundary. Novel numerical algorithms with theoretical guarantees are presented for efficiently computing recovery boundaries and safety margins.
Unlike prior methods, which tend to be overly conservative and restricted
to low dimensional parameter space, these methods compute safety margins to
arbitrary user-specified accuracy and do so efficiently in high dimensional parameter space.
% say something about how the algorithms work? ex. trajectory sensitivities?
The efficacy of the methods is demonstrated using the
IEEE 39-bus benchmark power system, where safety
margins are computed for cases that consider up to 86~parameters, and reveal
unexpected safety implications that would not have been observed otherwise.
\end{abstract}

%% \begin{IEEEkeywords}
%%   *** change !!! ***
%% power system dynamics, large disturbance stability.
%% \end{IEEEkeywords}

% Use this to place sponsorships
%\thanksto{\noindent Submitted to the 21st Power Systems Computation Conference (PSCC 2020).}

\section{Introduction}

Large disturbances, such as faults, are an inevitable part of power system operation, and in severe cases may lead to blackout conditions.
In order to ensure safe and reliable operation, it is
therefore important for power system operators to determine whether their
system is vulnerable to credible disturbances.
If the system is found to be vulnerable, i.e., the likelihood of recovery from potential disturbances is too small, 
system operators should take preventive action by manoeuvring the system to a
safer operating point, thereby reducing the vulnerability.

Whether the system recovers from a particular disturbance to a
desirable operating point depends upon many factors, including loading
pattern, controller settings, and load characteristics.  These factors
usually appear in power system models as ``known'' parameters, though
they are often not known with absolute certainty.  As a result, even
if vulnerability to a potential disturbance has been assessed for a
standard set of parameter values, those results may be unreliable and
misleading due to the uncertainty inherent in the parameters of the
actual system.  Therefore, we first introduce a notion of safety
margins for disturbance vulnerability assessment which quantitatively
establishes the smallest change in parameter values, from their
nominal values, that would result in the system
failing to recover from the
  disturbance to the desired operating point.

%% This work then develops novel, computationally efficient algorithms for
%% numerically computing safety margins in high dimensional parameter space,
%% which provides power system operators with a new tool to accurately assess
%% fault vulnerability while incorporating a large variety of relevant factors.

The system is not vulnerable to a potential disturbance if clearance
of that disturbance results in recovery to a desired operating
point. The ability of the system to recover depends on the system
parameter values.  Therefore, we define the {\em recovery region} for
a particular disturbance to be the set of parameter values for which
the system will recover to the desired operating point.  Furthermore,
we define the {\it recovery boundary} to be the boundary of the
recovery region in parameter space.  For a given nominal set of
parameter values, the {\it safety margin} is then defined as the
distance from that nominal case to the recovery boundary.  Thus, if
perturbations from the nominal parameter set are less than the safety
margin then the system is guaranteed to be able to recover from the
disturbance. If parameter uncertainty is constrained to the recovery
region then the system is guaranteed to recover from the specified
disturbance.

In the special case where the parameter space under consideration is chosen to
be the clearing time of a fault (and, thus, one dimensional), the recovery
boundary will consist of the critical clearing time
\cite[Chapter 9]{Sa97}, which is the maximum clearing time for which the
system is just marginally unable to recover from a particular fault,
and the safety margin will be the difference between the critical clearing
time and the nominal clearing time.
As the parameter space under consideration can be chosen to consist
of any desired collection of system parameters, the definition of the recovery
boundary can be thought of as a generalization of this notion of
critical clearing time to arbitrary selections of parameters,
thereby incorporating and quantifying the impact of parametric uncertainty
on disturbance vulnerability.

Typical industry practice involves the construction of conservative
approximations of the recovery boundary known as nomograms
\cite{Mc97}. Such methods are typically limited to two- or
three-dimensional parameter space. Several extensions to these
traditional techniques have been proposed \cite{Ma12}.  However, the
number of computationally-intensive dynamic simulations required by
these methods increases rapidly with parameter dimension.
Consequently, they are still limited to low dimensional parameter
spaces and are constructed using off-line processes, with limited
capability for updating to reflect real-time conditions. The small
number of parameters of interest are chosen based on operational
intuition and experience. Consequently, many parameters which could be
critical to system recovery may be overlooked. This is particularly
true given variability and uncertainty arising from renewable
generation, controller settings, and load characteristics.
Furthermore, existing methods are inherently approximate
and can be overly conservative. Hence, it is desirable to develop
techniques that offer more accurate and efficient computation of the
recovery boundary, allowing rapid assessment of safety margins in
higher dimensional parameter space.

%% \mfc{In fact, we do not have many modern references at all (excluding self
%%   references).  Do you know of any more modern references that we could
%%   include?}

Much of the classical work in rapid assessment of safety margins has
focused on developing estimates of the region of
attraction (RoA) of the desired operating point
\cite{Sa97,Ch15,Ch11,Tr96}.
However, as the exact identification of the RoA in high dimensional power
system models is typically intractable, existing methods are inherently
approximate and can be overly conservative.
Some approaches attempt to reduce conservatism using knowledge of a particular
equilibrium point, called the controlling unstable equilibrium point
(CUEP), with specific dynamical properties. However, locating the CUEP can be
challenging for high dimensional system models.
In addition, the majority of existing methods for estimating safety margins
focus on parameters which do not influence
post-disturbance dynamics, such as the fault clearing time, which excludes
many important system parameters.  Finally, it is often unclear how
to incorporate nonsmoothness, such as from saturation of
controllers like the AVR and PSS, which can have a
large influence on vulnerability.

Prior work \cite{Fi16,Fi18} developed efficient methods to numerically
compute the recovery boundary to arbitrary precision in
two-dimensional parameter space, and to determine safety margins in
arbitrary dimensional parameter space.  However, those methods require
prior knowledge of the CUEP of a disturbance.
Locating the CUEP
for a particular disturbance is challenging, particularly when considering
events such as controller saturation.
Other prior work \cite{Fi19b} required the numerical integration of
  additional high order dynamical systems, and was often inefficient or
  even intractable for high dimensional systems.
%\mfc{I wanted to mention our earlier work in \cite{Fi19b} because it has a
%  similar title and we might be asked to differentiate this paper from it.}

This paper develops novel, computationally efficient algorithms for
numerically computing the recovery boundary in two-dimensional parameter
space and determining safety margins in higher dimensional parameter space. These tools facilitate accurate, real-time assessment of disturbance vulnerability while capturing full nonlinear, non-smooth model details. The recovery boundary
and safety margins are computed to arbitrary precision, so they are not conservative. Furthermore, they can incorporate any number and type of parameters.
Crucially, these algorithms do not require prior knowledge of the CUEP, and
therefore can be directly and efficiently applied to realistic power systems.
%Furthermore, these methods can include nonsmooth limits, such as controller
%saturation limits, and therefore provide more accurate computations of the
%recovery boundary and safety margins than was previously possible.

Computation of points on the recovery boundary relies on two key theoretical insights.
%% First, the closest point on the recovery boundary in parameter space has
%% corresponding post-fault initial condition on the RoA boundary of the desired
%% equilibrium point in state space.
\begin{enumerate}
	\item As parameter values from within the recovery region approach the recovery
	boundary in parameter space, the post-disturbance initial condition in state-space, i.e., the system state at the instant the disturbance is cleared, approaches the region of attraction (RoA) boundary \cite{Sa97} of the desired equilibrium point.
	Therefore, to compute safety margins it suffices to find the closest
	parameter value to the given value whose corresponding post-disturbance initial
	condition lies on the RoA boundary in state space.
	\item Since this post-disturbance initial condition lies on the RoA boundary, its
	trajectory is infinitely sensitive to small perturbations. This is because incremental changes in parameter values could push the trajectory either inside or outside the RoA,
	leading to very different asymptotic behavior.
\end{enumerate}

Motivated by these ideas, the algorithms compute points on
the recovery boundary by varying
parameter values so as to maximize the trajectory sensitivities.
This is accomplished by numerically computing the trajectory sensitivities,
which are the derivatives of the trajectory with respect to parameter values. These sensitivities can be efficiently computed as a byproduct of numerically integrating the underlying system dynamics \cite{Hi00}. For implementation, however, it is numerically advantageous to minimize the inverse sensitivities rather than maximizing them directly. These ideas are applied to develop algorithms for numerically tracing
the recovery boundary in two-dimensional parameter space, and for finding safety margins in higher dimensional parameter space.

The algorithms are demonstrated on the IEEE 39-bus benchmark power system. They are used to compute safety margins in three cases that have parameter space dimensions of 38, 76 and~86, respectively. This investigation reveals unexpected dynamic behavior, with
important safety implications, which would not have been otherwise observed.

The paper is organized as follows.  Section~\ref{sec:theory} provides
the theoretical setting and justification for the algorithms, with the
details of the algorithms then given in
Section~\ref{sec:algo}. Section~\ref{sec:model} introduces the IEEE
39-bus benchmark power system test case, and Section~\ref{sec:res}
demonstrates the algorithm characteristics.  Section~\ref{sec:conc}
offers concluding remarks.

\section{Theory}\label{sec:theory}

\subsection{Model}

Power systems exhibit dynamic behavior which consists of periods of
smooth dynamics interspersed with discrete events, such as controllers
encountering saturation limits. Such behavior is commonly
modeled as a system of switched differential and algebraic equations
(switched DAEs):
\begin{align}
  \dot{x} &= f(x,y) \label{eq:dae1} \\
  0 &= g(x,y) \Lleftarrow \begin{cases}
    g_i^+(x,y), \quad s_i(x,y) > 0 \\
    g_i^-(x,y), \quad s_i(x,y) < 0,
    \end{cases} \label{eq:dae2}
\end{align}
where $g$ is composed of sets of switched algebraic equations $g_i^+,g_i^-:\mathbb{R}^{n+m} \to \mathbb{R}^{m_i}$, with $\sum_i m_i = m$, and the composition is determined by the signs of the corresponding switching indicator functions $s_i:\mathbb{R}^{n+m} \to \mathbb{R}$ whose zero level sets describe the switching surfaces, the dynamic states $x \in \mathbb{R}^n$ represent quantities such as rotor angles and frequencies whereas the algebraic states $y \in \mathbb{R}^m$ represents quantities such as bus voltages and line currents. Dynamics are driven by $f:\mathbb{R}^{n+m} \to \mathbb{R}^n$.
It is quite realistic to
assume that $f$, $g_i^+$, and $g_i^-$ are continuously differentiable. Note that switching occurs whenever $s_i$ passes through zero, i.e., a
switching surface is encountered, but otherwise, away from switching surfaces, the dynamics are smooth.
%\mfso{Furthermore, recent
%  theoretical work} \cite{Fi19}
%\mfso{has shown that the
%algorithms developed later in the paper
%extend to a large class of hybrid dynamical systems which permit
%switching behavior, such as controller limits, provided
%\ih{the {\em history} (time ordering) of
%switching events does not change}} \cite{piccoli98a}
%% \mf{the trajectory does not graze (i.e. lie tangent to) switching surfaces}.
%% \mfc{Local convergence guarantees do require a constant {\em history} near
%%   a particular critical parameter value, but the algorithms can still work
%%   nonlocally even when the switching history changes (which is what we observe
%%   in practice) provided that there is no grazing of switching surfaces
%%   \cite{Bu16}.}}
Such hybrid dynamical systems
accurately reflect realistic power system behavior.
%\mfso{
%For simplicity of presentation though, subsequent discussion will refer only to
%the DAE model \eqref{eq:dae1}-\eqref{eq:dae2}.}

The desired steady state operating point is given by a stable
equilibrium point (SEP) of \eqref{eq:dae1}-\eqref{eq:dae2} which has
an associated RoA\@.  Consider a particular disturbance, and let the
post-disturbance initial condition $x_0$ be the system state at the instant
when the disturbance clears.  Let $\mathbb{R}^P$ be a space of parameters of
the system, and let $p \in \mathbb{R}^P$ denote a vector of parameter
values.  As $p$ varies, so do the RoA and the post-disturbance initial
condition $x_0(p)$, with $x_0:\mathbb{R}^P \to \mathbb{R}^n$
capturing this parametric influence. Let $x \big(p,x_0(p),t \big)$
denote the dynamic states in the solution of
\eqref{eq:dae1}-\eqref{eq:dae2} at time~$t$ for parameter value $p$
and initial condition $x_0(p)$.

\subsection{Trajectory sensitivities}

Trajectory sensitivities are partial derivatives of the time varying
states with respect to initial conditions and/or parameters. It is shown in \cite{Hi00,Si19} that they can be
efficiently computed as a byproduct of the numerical integration of
the underlying dynamics for the switched DAEs of
  \eqref{eq:dae1}-\eqref{eq:dae2}.  For the vector
of parameters $p$, define the trajectory sensitivities $\chi$ of the
dynamic states $x$ with respect to $p$ as:
\begin{align}
\chi(p,t) := \frac{\partial x(p,x_0(p),t)}{\partial p},
\end{align}
i.e., the partial derivative of the dynamic states\footnote{The
  trajectory sensitivities $\frac{\partial y}{\partial p}$ of the
  algebraic states $y$ with respect to parameters $p$ can be similarly
  defined. Those sensitivities, however, are not used by the
  algorithms that are presented subsequently.} of the system at time
  $t$, starting from the post-disturbance initial condition $x_0(p)$.  These
  trajectory sensitivities measure the sensitivity of the system
  dynamic states along a trajectory to small changes in parameter
  value.

\begin{figure}
	\centering
	\includegraphics[width=0.5\textwidth]{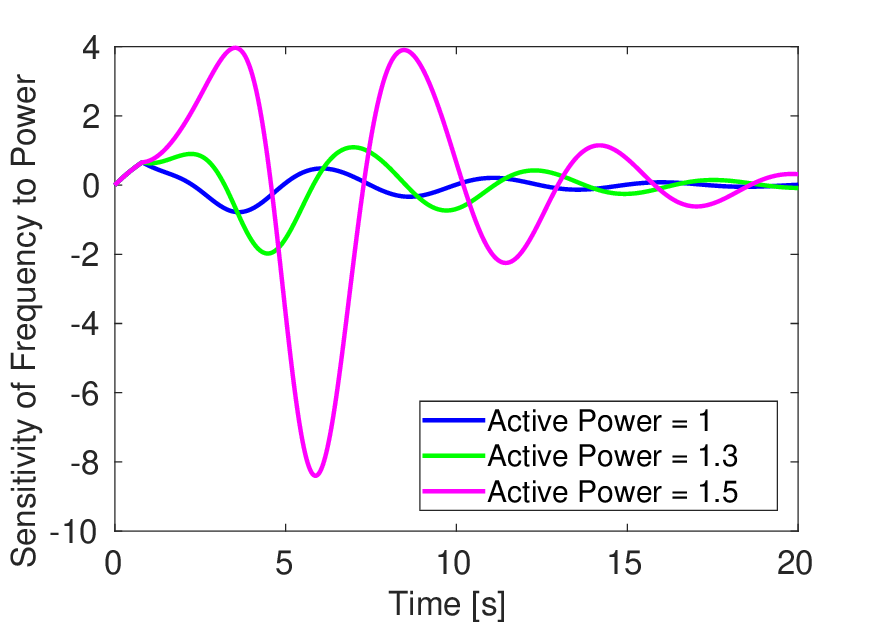}
	\caption{Sensitivity of the post-fault frequency to
		generator mechanical power, as the generator
		power approaches the recovery boundary.}
	\label{fig:classical}
\end{figure}

\medskip
\noindent {\bf Example.} To develop intuition for the behavior of trajectory sensitivities for
parameter values near the recovery boundary, consider the simple example of
a single machine infinite bus with classical machine dynamics \cite{Sa97}.
The disturbance considered in this system is a fault which is
modeled by setting the electrical torque to
zero until the fault is cleared.  Fig.~\ref{fig:classical} shows the
sensitivity of the frequency to generator mechanical power as that
power value approaches the recovery boundary.  Observe that trajectory
sensitivities grow considerably larger as the recovery boundary is approached. \hfill$\Box$

\medskip
More generally, the intuition behind this observation is that at the
recovery boundary, small perturbations in parameter values could
  push the trajectory to either side of the RoA boundary, i.e., either
  recovery from the disturbance or
  failure to recover.
  Hence, the trajectory
  becomes infinitely sensitive to small changes in parameter values.

\subsection{Inverse sensitivities}

To generalize the above intuition, define the scalar function
$G:\mathbb{R}^P \to \mathbb{R}$ by:
\begin{align}
 H(p,t) &= \frac{1}{||\chi(p,t)||_1}, \label{eq:H} \\
G(p) &= \inf_{t \geq 0} H(p,t), \label{eq:G}
\end{align}
where the matrix norm is
defined by $||M||_1 = \sum_{i,j} |M_{ij}|$ for any matrix $M$. Hence,
$G(p)$ represents the minimum over time of the inverse of the norm of
the trajectory sensitivities. It follows that trajectory sensitivities
diverging to infinity is equivalent to $G$ approaching zero.  Under
practical assumptions, for a large class of nonlinear systems, the
following result has been proven in
\cite[Theorem~1]{Fi25}:

%% \medskip
%% \noindent {\bf Theorem 1.}  The function $G$ is well-defined,
%% continuous, and strictly positive over the recovery region, the
%% recovery boundary is equal to $G^{-1}(0)$, and for any convergent
%% sequence of parameter values $\{p_k\}_{k=1}^\infty$ in the recovery
%% region, if $\lim_{k \to \infty} G(p_k) = 0$ then $\lim_{k \to \infty}
%% p_k$ lies on the recovery boundary. \hfill$\Box$

\medskip
\noindent {\bf Theorem 1.}  The function $G$ is well-defined,
continuous, and strictly positive over the recovery region,
for $p^*$ in the recovery boundary almost everywhere $G(p^*) = 0$,
%the recovery boundary is equal to $G^{-1}(0)$,
and for any convergent
sequence of parameter values $\{p_k\}_{k=1}^\infty$ in the recovery
region, if $\lim_{k \to \infty} G(p_k) = 0$ then $\lim_{k \to \infty}
p_k$ lies on the recovery boundary. \hfill$\Box$

\medskip
In particular, Theorem 1 implies that solving for $p$ such
that $G(p) = 0$ will drive $p$ onto the recovery boundary.
Furthermore, Theorem 1 also implies that, up to a set of measure zero,
  the recovery boundary is equal to $G^{-1}(0)$.
These motivate
the development of algorithms for computing points on the recovery boundary.

\medskip

In order to develop algorithms for solving $G(p) = 0$, we require the derivative $DG(p)$. Note that,
\begin{equation}
  DG(p) = \begin{bmatrix}
    \pd{G}{p_1}(p) & \pd{G}{p_2}(p) & \hdots & \pd{G}{p_P} \label{eq:dg}
  \end{bmatrix}.
\end{equation}
For any parameter $p$ that lies inside the recovery region, by
the proof of \cite[Theorem 1]{Fi25} there exists a time
$\hat{t}(p)$ such that,
\begin{equation}
  G(p) = \min_{t \geq 0} H(p,t) = H(p,\hat{t}(p)). \label{eq:Gp}
\end{equation}
Since $\hat{t}(p)$ is the point in time at which $H(p,t)$ achieves a minimum,
it is an extremal point of $H(p,t)$ which implies,
\begin{equation}
\pd{H}{t}(p,\hat{t}(p)) = 0. \label{eq:zero}
\end{equation}
From \eqref{eq:Gp},
\begin{align}
  \pd{G}{p_j}(p) &=
  \pd{}{p_j} %\ih{H(p,t)|_{(p,\hat{t}(p))}}
  \Big( H(p,\hat{t}(p)) \Big)
  \nonumber \\
  &= \pd{H}{p_j}(p,\hat{t}(p))
  + \pd{H}{t}(p,\hat{t}(p)) \pd{\hat{t}}{p_j}(p) \nonumber \\
  &= \pd{H}{p_j}(p,\hat{t}(p)), \label{eq:dGdp}
\end{align}
where the last step follows by substituting in \eqref{eq:zero}.
So, to compute $DG$ it suffices to compute $\pd{H}{p_j}$.
To do so, define the following notation for $i,j \in \{1,2,...,P\}$:
\begin{align*}
  \chi_{[i]}(p,t) &:= \pd{x}{p_i}(p,x_0(p),t), \\
  \chi_{[ij]}(p,t) &:= \pdd{x}{p_i}{p_j}(p,x_0(p),t),
\end{align*}
where $\chi_{[i]}(p,t)$ and $\chi_{[ij]}(p,t)$ are both
$n$-dimensional vectors\footnote{Note that $\chi_{[ij]}(p,t)$ differs
  from standard notation where $\chi_{ij}(p,t) \equiv
  \pd{x_i}{p_j}(p,x_0(p),t)$.}.  Then $\pd{H}{p_j}$ is given by,
\begin{align}
  \pd{H}{p_j}(p,t) =
  -\f{\sum_{i=1}^P \text{sign}(\chi_{[i]}(p,t))^\intercal \chi_{[ij]}(p,t)}
  {||\chi(p,t)||_1^2}, \label{eq:Hp}
\end{align}
where for any vector $v$, $(\text{sign}(v))_i = 1$ if $v_i \geq 0$ and
$(\text{sign}(v))_i = -1$ if $v_i < 0$. As mentioned earlier, first-order
trajectory sensitivities $\chi_{[i]}$ can be computed efficiently as a byproduct of the
underlying numerical integration. This is also the case for second-order sensitivities $\chi_{[ij]}$, as shown in \cite{Si19,Fi19b}. Combining
\eqref{eq:dg}, \eqref{eq:dGdp} and \eqref{eq:Hp} therefore enables
efficient computation of $DG(p)$.

\section{Algorithms}\label{sec:algo}

Based on Theorem 1, the recovery boundary consists precisely of the
parameter values $p$ such that $G(p) = 0$. This section formulates
algorithms for, 1) finding a
point on the recovery boundary in the case of one-dimensional
parameter space, 2) numerically tracing the recovery boundary in two-dimensional parameter space, and 3) finding the closest point on the
recovery boundary in higher dimensional parameter space.

\subsection{One-dimensional parameter space}\label{sec:one}

To find a point on the recovery boundary, the goal is to solve,
\begin{equation}
  G(p) = 0. \label{eq:goal}
\end{equation}
Earlier theoretical work \cite{Ro12} and numerical experiments
\cite{Ng02} suggest that $G$ is approximately affine for parameter
values near the recovery boundary.  Motivated by this linear structure, we
solve \eqref{eq:goal} using Newton-Raphson with backtracking, which is
an iterative algorithm that converges rapidly for affine equations.

Let $k$ be the latest iteration such that $p^k$ lies in the
  recovery region. (At initialization, $k=0$ with $p^0$ user
  specified.) The subsequent iteration $i=k+1$ computes the updated
  parameter value $p^i$ according to the Newton-Raphson step,
\begin{equation}
	p^{i} = p^k - \mu G(p^k)/DG(p^k), \label{eq:update}
\end{equation}
where $\mu = \big(\frac{1}{2}\big)^{i-k-1}$. A full time-domain
simulation is performed using parameter values $p^i$ to determine
whether the system recovers to the SEP or not, hence establishing
whether $p^i$ lies within or outside the recovery region. The values
for $G(p^i)$ and $DG(p^i)$ are also computed. If $p^i$ lies within the
recovery region then $k$ is updated by setting $k=i$ and $i$ is
reinitialized to $i=k+1$. Otherwise, $k$ remains unchanged and $i$ is
incremented, $i=i+1$. The process then repeats.

The algorithm ensures that $\mu=1$ for the first step beyond iteration
$k$, so the full Newton-Raphson update is considered. However, if that
results in progressing beyond the recovery region then the step length
$\mu$ is halved. This achieves a backtracking line search along the
Newton-Raphson direction until a parameter value inside the recovery
region is obtained.

%At each iteration $s$, we perform a full time-domain simulation to
%evaluate $G$ and its derivative $DG$.  The results of that simulation
%also reveal whether $p^s$ lies inside or outside the recovery region, based on
%whether the system recovers or does not recover to the desired SEP\@.
%Let $k$ be the largest iteration with $k \leq s$ such that $p^k$ lies
%in the recovery region.  In other words, $k$ represents the most recent
%iteration
%at which $p^k$ lies in the recovery region.  Let
%$\mu = \left(\frac{1}{2}\right)^{s-k}$ denote the step length.  We perform
%the update
%\begin{align}
%p^{s+1} = p^k - \mu ~G(p^k)/DG(p^k). \label{eq:update2}
%\end{align}
%If $p^s$ lies inside the recovery region, so that $k = s$, then $\mu = 1$
%and this reduces to the standard Newton-Raphson update.
%On the other hand, if $p^s$ lies outside the recovery region, then we start
%from the most
%recent value $p^k$ that lies inside the recovery region,
%determine the direction of its
%Newton-Raphson update, and do a backtracking line search along that direction
%until a parameter value inside the recovery region is reached.
%In particular, the backtracking is performed by reducing the distance traveled
%along the Newton-Raphson search direction by a factor of two at every iteration
%until the recovery region is encountered.

Performing the update of \eqref{eq:update} requires computation of $G$
and its derivative $DG$, which proceeds as follows.
During a time-domain simulation corresponding to parameter value $p^k$,
the time $\hat{t}(p^k)$ at which $H(p^k,t)$ achieves its minimum (over time) is observed.
Then $G(p^k) = H(p^k,\hat{t}(p^k))$, and $DG(p^k)$ is computed
using \eqref{eq:dg}, \eqref{eq:dGdp} and \eqref{eq:Hp} with $P = 1$.
These are then used to perform the update of \eqref{eq:update} to
compute $p^i$.

This process is repeated iteratively until $G(p^k)$ converges towards
zero, which causes $p^k$ to converge to the recovery boundary. This
outcome is guaranteed by the following theorem, which was proven in
\cite[Theorem 2]{Fi25} under similar assumptions as
in Section~\ref{sec:theory}:

\medskip
\noindent {\bf Theorem 2}. For $p^0$ sufficiently close to the recovery
boundary,
the sequence $\{p^k\}_{k=0}^\infty$ generated by \eqref{eq:update}
is well-defined and will converge
to a unique parameter value on the recovery boundary.
\hfill$\Box$

\medskip
As with any numerical computation, satisfaction of \eqref{eq:goal} can only
be achieved up to a small convergence tolerance $\epsilon > 0$.
In other words, it is only
possible to numerically compute $p^*$ such that $|G(p^*)| \leq \epsilon$,
rather than realizing $G(p^*) = 0$ exactly.
The same is true for the algorithms presented in Sections~\ref{sec:two}
and \ref{sec:high}, where $\epsilon > 0$ will be the tolerance chosen
such that the solution $p^*$ satisfies $|G(p^*)| \leq \epsilon$.

\subsection{Two-dimensional parameter space}\label{sec:two}

\begin{figure}
  \centering
  \includegraphics[width=0.5\textwidth]{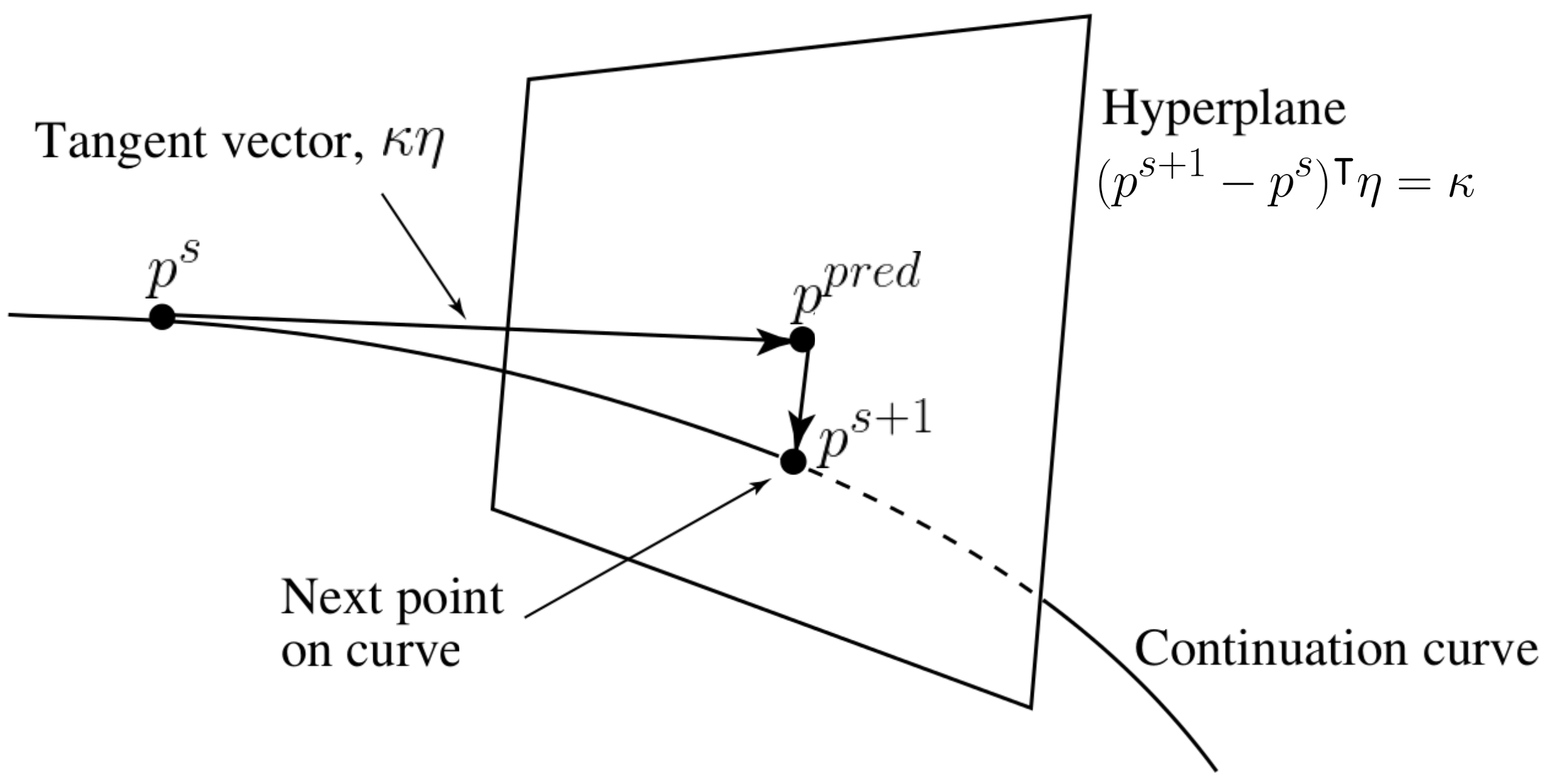}
  \caption{The predictor-corrector continuation process for iteratively
  tracing a sequence of points along a curve.} \label{fig:predcor}
\end{figure}

By Theorem 1, the recovery boundary is equal to $G^{-1}(0)$.  In two-dimensional parameter space, $G:\mathbb{R}^2 \to \mathbb{R}$ is a
scalar equation with two free variables (parameters). Because the number of free variables exceeds the number of equations by one, the recovery boundary is typically a one-dimensional curve. The goal is to numerically trace this curve $G^{-1}(0)$ by
iteratively computing a sequence of points that lie along the curve. This is accomplished
using the following continuation method, which alternates between a
predictor step and a corrector step, as shown in
Fig.~\ref{fig:predcor}.  Let $s$ denote the current point on the
  curve, and $p^s$ be the corresponding parameter values.  Then $p^s$
  must satisfy $G(p^s) = 0$.

The predictor step generates a first order prediction of the next
point on the curve, and proceeds as follows.  First, we obtain the
unit tangent vector to the curve $G^{-1}(0)$ at $p^s$.  This
can be done by noting that $DG(p^s)$ is orthogonal\footnote{Because $G$ is constant along the curve, its derivative $DG$ must be orthogonal to it.} to the curve $G^{-1}(0)$.  Hence, any vector orthogonal to $DG(p^s)$ must be tangent to $G^{-1}(0)$ since the parameter
space is exactly two-dimensional.  We can compute $DG(p^s)$ at each iteration
using \eqref{eq:dg}, \eqref{eq:dGdp} and \eqref{eq:Hp} for $P = 2$.
Then, writing $DG(p^s) = [a~b]$, a tangent vector to the curve is
given by $[b~- \! a]^\intercal$.  Let $\eta$ denote $[b~- \! a]^\intercal$ divided by its
norm, so that $\eta$ is the unit vector tangent to the curve.  Let
$\kappa \in \mathbb{R}$ be the step size for the prediction. Therefore the predicted point is given by,
\begin{align}
p^{pred} = p^s + \kappa \eta. \label{eq:pred}
\end{align}

Once a predicted point has been obtained, the next step is to correct back onto the curve $G^{-1}(0)$. To do so, an orthogonal projection from the predicted point onto the curve is performed.  In
particular, the next point on the curve $p^{s+1}$ is chosen such that
the vector $(p^{s+1}-p^{pred})$ is orthogonal to the vector
$(p^{pred}-p^s)$, so that,
\begin{align*}
(p^{s+1}-p^{pred})^\intercal(p^{pred}-p^s) = 0.
\end{align*}
Using \eqref{eq:pred} and the fact that $\eta$ is a unit vector, so
$\eta^\intercal\eta = 1$, the above constraint simplifies to,
\begin{align}
  (p^{s+1}-p^s)^\intercal\eta - \kappa = 0. \label{eq:hyper}
\end{align}
Therefore, correction back onto the curve requires solution of the system of
equations:
\begin{align}
  F(p^{s+1}) := \begin{bmatrix}
    G(p^{s+1}) \\
    (p^{s+1}-p^s)^\intercal\eta - \kappa
  \end{bmatrix} = 0.
\end{align}
For notational convenience, replace $p^{s+1}$ by $\wp$. Solution
  of $F(p^{s+1}) \equiv F(\wp) = 0$, which consists of two equations
  with two free variables (parameters), can again be achieved using
  Newton-Raphson with backtracking:
\begin{align}
  \wp^i = \wp^k - \mu DF(\wp^k)^{-1}F(\wp^k) \label{eq:predcor}
\end{align}
where  $\mu = \left(\f{1}{2}\right)^{i-k-1}$, iteration indices $i,k$ are
updated iteratively as in \eqref{eq:update} and,
\[
  DF(\wp) = \begin{bmatrix}
    DG(\wp) \\
    \eta^\intercal
    \end{bmatrix}.
\]
At each iteration, if $p^{pred}$ lies in the recovery region then
  set $\wp^1 = p^{pred}$.
  %the continuation process can proceed as normal.
  Otherwise, if
  $p^{pred}$ does not lie in the recovery region, then we perform a
  line search along the line \eqref{eq:hyper} using, for example, the
  bisection or golden section search methods, to find a parameter
  value on \eqref{eq:hyper} that lies in the recovery region, and set
  $\wp^1$ equal to that parameter value.
  Then, we proceed with the continuation process using the update
  \eqref{eq:predcor} at each iteration.
Upon convergence, Newton-Raphson will
provide $\wp \equiv p^{s+1}$ such that $F(p^{s+1}) = 0$, which ensures
that $G(p^{s+1}) = 0$, so $p^{s+1}$ is on the curve $G^{-1}(0)$. This
is guaranteed by the following theorem, which was proven in
\cite[Theorem~3]{Fi25} under similar assumptions as
in Section~\ref{sec:theory}:

\medskip
\noindent {\bf Theorem 3}. For $p^s$ satisfying $G(p^s) = 0$ and
$\kappa > 0$ sufficiently small almost everywhere, and for
$\wp^1$ contained in the hyperplane \eqref{eq:hyper} and in the recovery
region, the sequence
$\{\wp^k\}_{k=1}^\infty$ generated by \eqref{eq:predcor} is well-defined
and will converge to a unique $p^{s+1}$ that lies in the hyperplane
\eqref{eq:hyper} and on the recovery boundary. \hfill$\Box$

\medskip
The prediction and correction steps then alternate until the recovery boundary has been traced.

\subsection{Higher dimensional parameter space} \label{sec:high}

Now suppose that $P>2$, so that $G^{-1}(0)$ may be more than
two-dimensional. (Recall that $G^{-1}(0)$ typically has dimension equal to the number of free variables (parameters) minus the number of constraints, in this case one scalar equation.)
Therefore, rather than tracing the recovery boundary, in this section our
goal will be to find the closest point on the multi-dimensional recovery boundary to an initial parameter value. This will provide a quantitative measure of the margin for safe operation in the sense that it provides the smallest change in parameter values that would lead to a failure to recover from the
disturbance.
%\mfso{As in Section~\ref{sec:two},
%as numerical precision is finite, instead of finding the closest point
%on $G^{-1}(0)$, we instead find the closest point on $G^{-1}(\epsilon)$ for some
%small $\epsilon > 0$.  This approximates the true recovery boundary to
%arbitrary precision, with the tolerance level set by $\epsilon$.}
%Let $A$ be any square positive semidefinite matrix which we call a weighting
%matrix.

Let $p_0$ be any given initial parameter value, such as the current system
conditions.
Our objective is to find $p \in G^{-1}(0)$ of minimum distance from
$p_0$, where we consider Euclidean distance (2-norm)\footnote{Note that Theorem~4, as proven in \cite{Fi25}, holds for more general norms of the form
  $(p-p_0)^\intercal A (p-p_0)$, where $A$ is symmetric positive definite.}
$||p-p_0||_2$.
This can be formulated as the optimization problem:
\begin{align}
  \begin{split}
&  \min_{p \in \mathbb{R}^P} \f{1}{2}(p-p_0)^\intercal(p-p_0) \\
    &\quad \text{s.t. } G(p) = 0, 
    \end{split} \label{eq:nonconvex}
\end{align}
where $(p-p_0)^\intercal(p-p_0) = ||p-p_0||_2^2$.
In general $G$ is nonconvex, so \eqref{eq:nonconvex} is a nonconvex
optimization problem that is challenging to solve.
However, historical work \cite{Ro12,Ng02}, as well as our own numerical experiments (see Fig.~\ref{fig:single}), suggest
that $G$ is approximately affine near the recovery boundary. Therefore,
we approximate $G(p)$ locally by its linearization and solve
a sequence of quadratic programs \cite{nocedal06a}.
In particular, at each iteration $k$ we replace the nonlinear constraint
$G(p^{k+1}) = 0$ with the corresponding affine constraint obtained by linearizing $G$
at $p^k$:
\begin{align*}
  DG(p^k)(p^{k+1}-p^k) + G(p^k) = 0.
\end{align*}
This gives the quadratic program,
\begin{align}
  \begin{split}
  &  \min_{p^{k+1} \in \mathbb{R}^P} \f{1}{2}(p^{k+1}-p_0)^\intercal(p^{k+1}-p_0) \\
  &\quad \text{s.t. } DG(p^k)(p^{k+1}-p^k) + G(p^k) = 0.
  \end{split} \label{eq:quad}
\end{align}
The Lagrangian for this problem is given by,
\begin{align*}
  \mathcal{L}(p^{k+1},\lambda^{k+1})
  &= \f{1}{2}(p^{k+1}-p_0)^\intercal(p^{k+1}-p_0) \\
  &\quad+ \lambda^{k+1} \big( DG(p^k)(p^{k+1}-p^k) + G(p^k) \big),
\end{align*}
where $\lambda^{k+1}$ is a scalar Lagrange multiplier.
The KKT conditions for an optimal solution yield,
\begin{align*}
  0 = \nabla \mathcal{L}(p^{k+1},\lambda^{k+1})
  = \begin{bmatrix}
    p^{k+1}-p_0 + DG(p^k)^\intercal\lambda^{k+1} \\
    DG(p^k)(p^{k+1}-p^k) + G(p^k)
  \end{bmatrix}.
\end{align*}
Rearranging, the solution is given by the linear system,
\begin{align*}
  \begin{bmatrix}
    I & DG(p^k)^\intercal \\
    DG(p^k) & 0
  \end{bmatrix}
  \begin{bmatrix}
    p^{k+1} \\
    \lambda^{k+1}
  \end{bmatrix}
  = \begin{bmatrix}
    p_0 \\
    DG(p^k)p^k - G(p^k)
    \end{bmatrix},
\end{align*}
which can be efficiently solved numerically, with $G(p^k)$ and $DG(p^k)$ computed using \eqref{eq:dg}, \eqref{eq:dGdp} and \eqref{eq:Hp} following
numerical integration of the first and second order trajectory sensitivities.

If $p^{k+1}$ lies outside of the recovery region, which is determined as a
byproduct of the simulation required to compute $G(p^{k+1})$ and
$DG(p^{k+1})$, then we instead move to the point $p^k + \mu(p^{k+1}-p^k)$,
where $\mu \in (0,1)$ is determined by a backtracking line search that
reduces by a factor of two at each iteration until the recovery region is
encountered, as in Section~\ref{sec:one}.

Sequential quadratic programming is repeated iteratively until the
sequence of parameters converges to the optimal solution of
\eqref{eq:nonconvex}. This outcome is guaranteed by
the following theorem, which was proven in \cite[Theorems~4-5]{Fi25} under
similar assumptions as in Section~\ref{sec:theory}:

\medskip
\noindent {\bf Theorem 4}. For $p_0$ sufficiently close to $\partial R$
almost everywhere, there exists a unique solution $p^*$ to \eqref{eq:nonconvex}.
Furthermore, the sequence $\{p^k\}_{k=1}^\infty$ starting from $p^1 = p_0$ and
generated by solving \eqref{eq:quad} and using a backtracking line search
converges to $p^*$.

%% \mfc{I think this theorem still works as written, provided that we interpret
%%   the solution of \eqref{eq:nonconvex} as a numerical optimization problem
%%   with tolerance $\epsilon$ on the equality constraints.}
%% \ihc{This sounds sensible though I can't really comment on the technical
%%   details.}
%% \mfc{Upon reflection, for numerical solvers \eqref{eq:nonconvex} is equivalent
%%   to allowing an $\epsilon$ tolerance around the equality constraint (for which
%%   case the theorem has been proven in the other paper), so I
%%   agree with my original comment.  I do not think any change is needed, but
%%   we could consider a footnote if desired.}
%% \ihc{I'm happy not saying anything.}

\section{Example}\label{sec:model}

\begin{figure}
  \centering
  \includegraphics[width=0.45\textwidth]{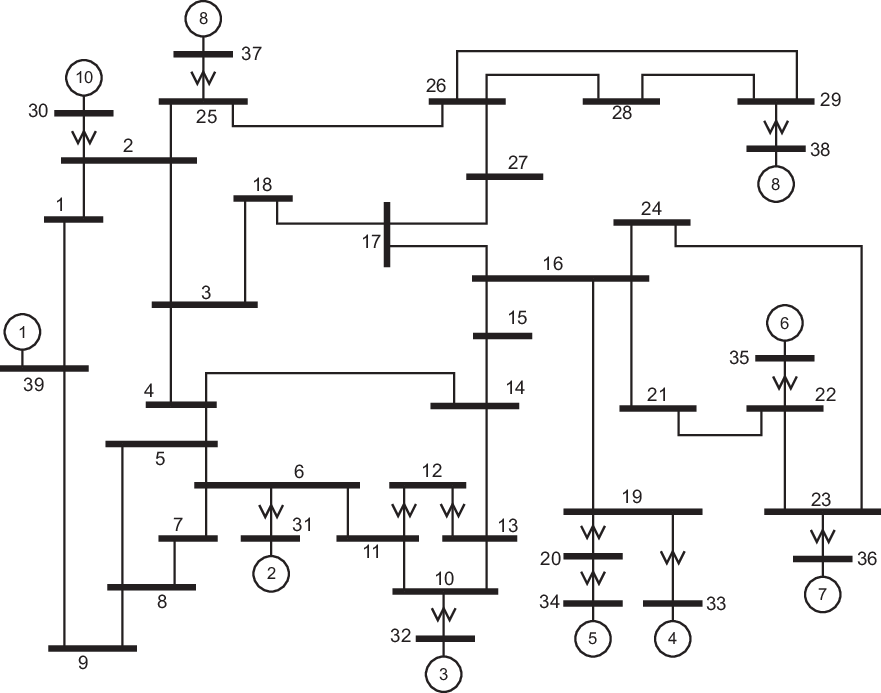}
  \caption{IEEE 39-bus benchmark power system.}
  \label{fig:net}
\end{figure}

The IEEE 39-bus benchmark test case \cite{Ps15}, shown in
Fig.~\ref{fig:net}, will be used to illustrate the algorithms of
Section~\ref{sec:algo}.  Generators are modeled using the 4th-order
machine model from \cite{Sa97}, while AVRs and PSSs are modeled
according to the IEEE standard \cite{ieee16a} ST1C and PSS1A models,
respectively. The full set of dynamic equations and system parameters
are given in \cite{Ps15}.  It is important to note that the nonsmooth
AVR/PSS controller limits are included in this system model, which is
vitally important for meaningful vulnerability assessment.  The
network is subject to a three-phase fault at bus 16, with fault
clearing at 0.2~s.  The fault is modeled as a switched shunt
reactance, $X_{fault} = 0.001$~p.u.

Many model parameters of the system are of interest for recovery
considerations.  A load scaling factor is introduced
which multiplies the active and reactive power loads at
every bus in the network.  As load is often
uncertain, it is a natural choice for assessing system recoverability.
An AVR gain scaling factor multiplies the AVR gain for every generator, and helps
to capture the impact of controller tuning on system stability.  The active and reactive power loads are represented by the standard
exponential form of voltage dependent load model. The voltage exponents for all active-power loads are set equal, likewise for all reactive-power loads. As load dynamics are notoriously difficult to model, these parameters serve to capture the impact of uncertain
load behavior on system recoverability.

\section{Results}\label{sec:res}

The algorithms of Section~\ref{sec:algo} were demonstrated on the benchmark test
case of Section~\ref{sec:model}.  The dynamic states used in the formation of the function $H$ in \eqref{eq:H}, and hence $G$ in \eqref{eq:G}, were restricted to the generator dynamic states
to avoid the possibility of the sensitivities of controller states dominating the sensitivities of the physical generator states.

\begin{figure}
  \centering
  \includegraphics[width=0.5\textwidth]{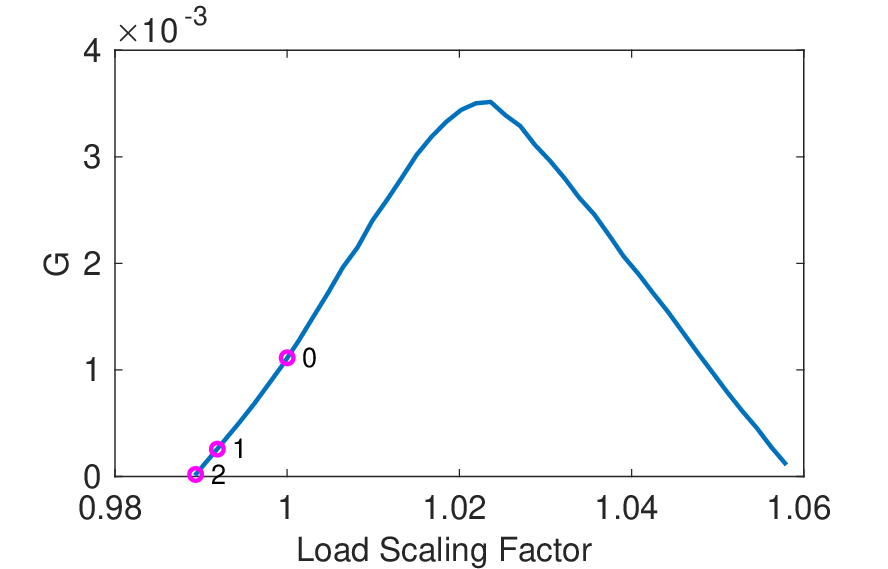}
  \caption{$G$ as a function of load scaling factor (blue line).
    The iterations of the one-dimensional parameter space algorithm are identified by circles and the iterations are labeled.} \label{fig:single}
\end{figure}

The algorithm for one-dimensional parameter space, which
finds a point on the recovery boundary, was applied with the parameter of interest being the load scaling factor.  Fig.~\ref{fig:single} shows that the
algorithm converged to the recovery boundary in just 2 iterations (3
simulations total).  Furthermore, it shows that $G$ is approximately
linear near the recovery boundary, both when the scaling factor is low (near 0.99) and high (near 1.06). The algorithm rapidly and accurately determines points on the recovery boundary in one-dimensional parameter space.

\begin{figure}
  \centering
  \includegraphics[width=0.5\textwidth]{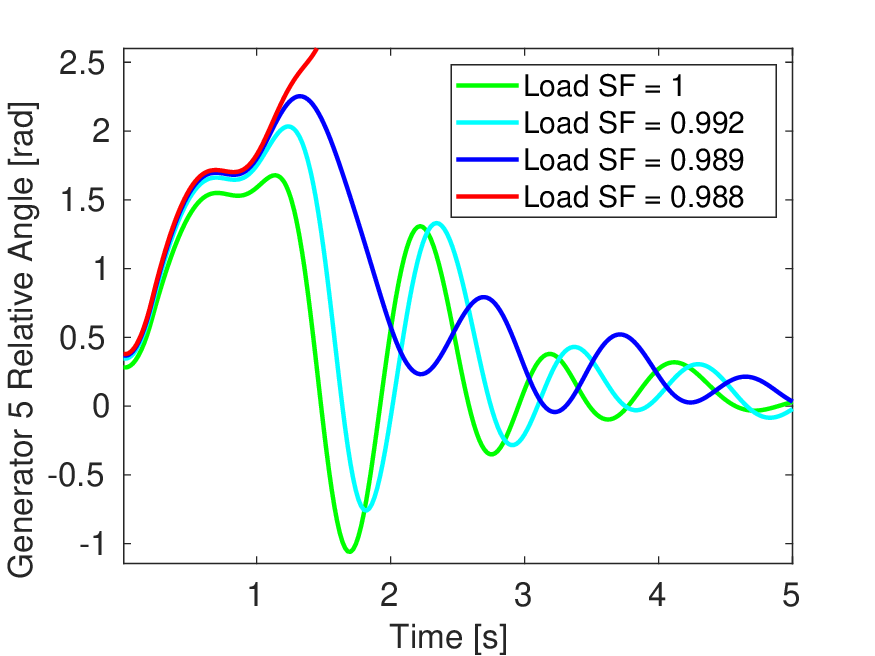}
  %% \caption{Frequency of generator~2 as a function of time for the
  %%   values of the background load scaling factor that correspond to
  %%   the iterations (inside the recovery region) of the one dimensional
  %%   algorithm.}
  \caption{The angle of generator 5 relative to generator 2 for the
    load scaling factor attained at each iteration of the
    one-dimensional algorithm. These values are shown in
    Fig.~\ref{fig:single}.} \label{fig:load_response}
\end{figure}

To observe the progression towards instability, note that generator~5  is the first generator to go unstable under stressed conditions. Fig.~\ref{fig:load_response} shows the relative angle of generator 5 as a
function of time for the values of the load scaling factor that were attained at each iteration of the algorithm, i.e., the points shown in Fig.~\ref{fig:single}. Observe that as the
load scaling factor approaches the recovery boundary, the initial fluctuations in the
relative angle of generator 5 grow larger, indicative of proximity to
instability.

\begin{figure}
  \centering
  \includegraphics[width=0.5\textwidth]{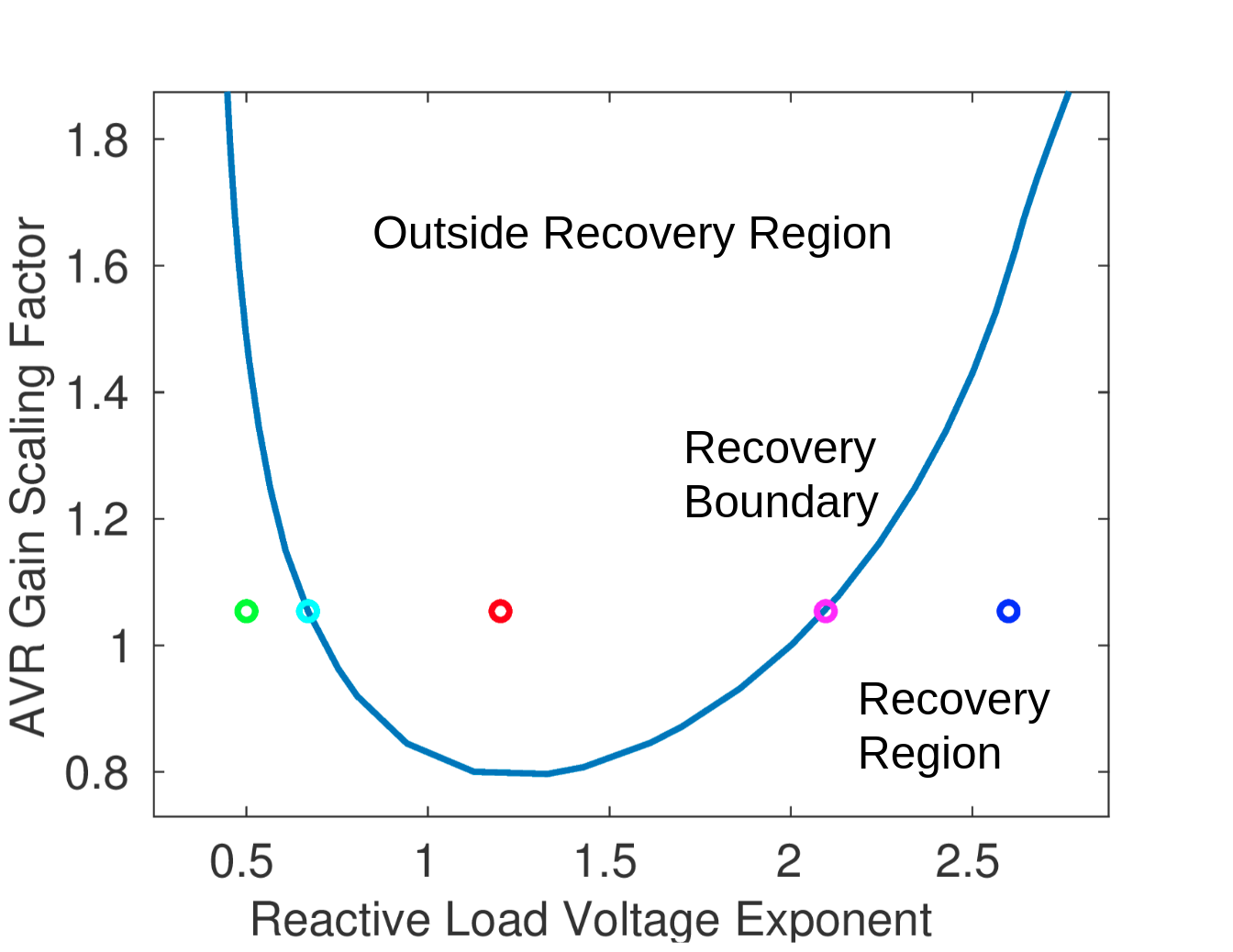}
  \caption{The recovery boundary and recovery region in the two-dimensional parameter space of AVR gain scaling factor and reactive load voltage exponent. Colored circles indicate the parameter values whose corresponding dynamic behavior is shown in Fig.~\ref{fig:exp_avr_response}.} \label{fig:cont}
\end{figure}

Once a point on the recovery boundary has been identified, the continuation
method described in Section~\ref{sec:algo} can be applied to
numerically trace the recovery boundary in two-dimensional parameter space.
This algorithm was applied for the two-dimensional parameter space
consisting of the AVR gain scaling factor and the reactive load voltage exponent. The
tolerance was set to $\epsilon = 10^{-5}$.  Fig.~\ref{fig:cont} shows
the recovery boundary, and hence recovery region, in this parameter space.
If the AVR gain scaling factor is reduced to around 0.8, i.e., gains are reduced by around 20\%, the system will recover regardless of the reactive load exponent. However, for higher values of the AVR gain scaling factor, a nonempty region outside the recovery region emerges.
For such cases, a wide range of reactive load voltage exponent values lie outside the recovery region. For example, if reactive loads exhibit constant current to constant voltage characteristics (voltage exponents in the range~1 to~2, respectively) and AVR gains are around their nominal values (scaling factor close to~1) then the system will not be able to recover from the fault.
%In particular, as the loads approach constant current and constant
%impedance loads, for nominal and slightly above nominal AVR gains the
%system will not be able to recover from the fault. 
In contrast, as reactive loads approach constant power characteristics (voltage exponent around~0), the system becomes more resilient and is able to recover from the fault. These observations run counter to standard intuition, which suggests that constant power loads have a more detrimental impact on system stability. This counter-intuitive behavior serves as an example of the potential offered by these algorithms for revealing dynamic behavior which would not otherwise have been anticipated.

\begin{figure}
	\centering
	\includegraphics[width=0.5\textwidth]{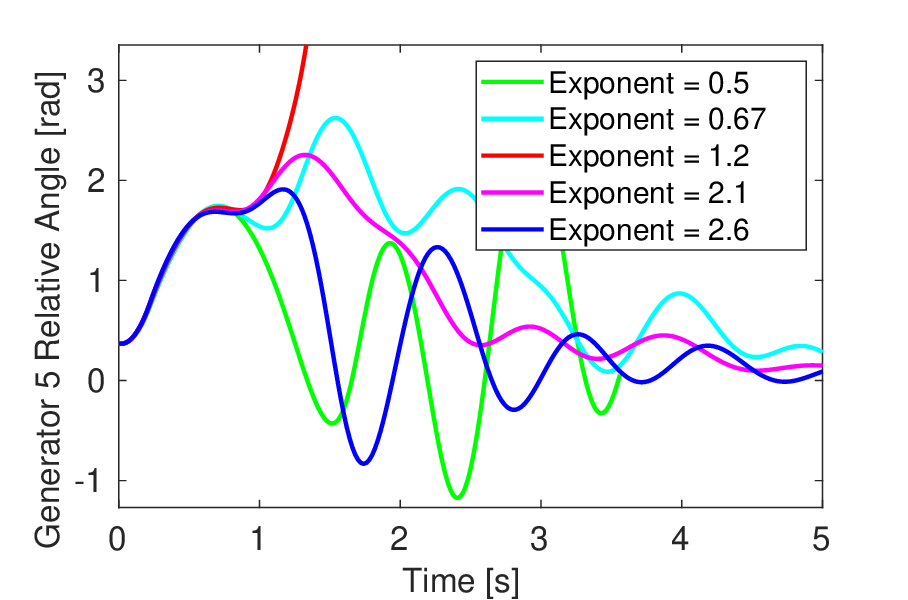}
	\caption{The angle of generator~5 relative to generator~2, for an AVR gain scaling factor of $1.054$ and the reactive load voltage exponents identified in Fig.~\ref{fig:cont}.} \label{fig:exp_avr_response}
\end{figure}

To observe the influence of reactive load voltage
exponents on system recovery, recall that generator~5 is the first
generator to go unstable as a result of the fault.
Fig.~\ref{fig:exp_avr_response} shows
the relative angle of generator~5 as a function of time for a fixed
AVR gain scaling factor of 1.054 and for several reactive load voltage exponent values identified
in Fig.~\ref{fig:cont}. For
sufficiently high or low exponents, the angle shows
smaller initial fluctuations and a more rapid return to synchrony. For
exponents near the recovery boundary, initial fluctuations are larger
and take a longer time to damp before resynchronization. For exponent
values between these recovery boundary values, the system loses synchronism and is unable to recover from the fault.  Hence, for intermediate voltage exponents, angle fluctuations exhibit instability, whereas high or low load voltage exponents result in recovery from the disturbance.

\begin{figure}
  \centering
  \includegraphics[width=0.5\textwidth]{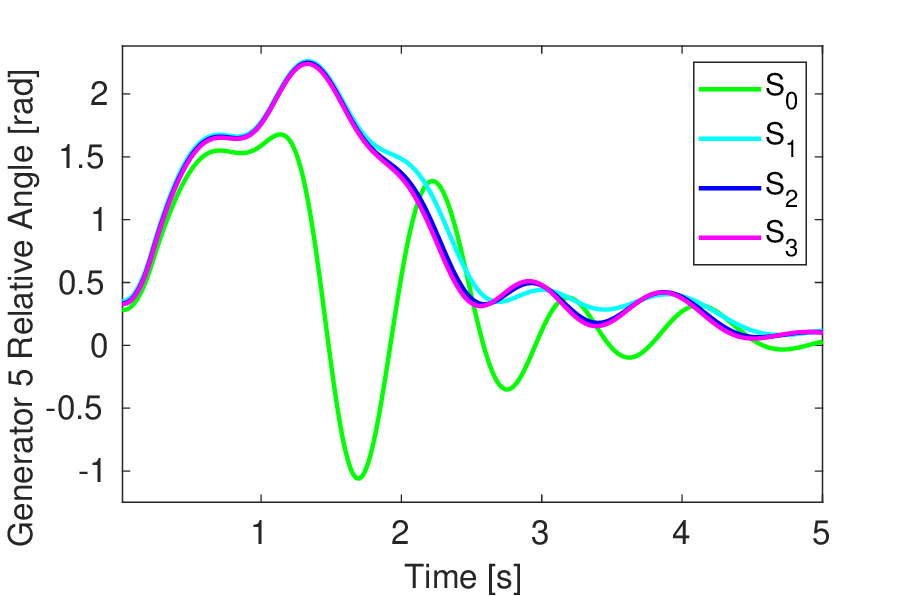}
  \caption{The angle of generator~5 relative to generator~2 for the solutions of the optimization algorithm for parameter sets $S_1$, $S_2$ and $S_3$, as well as for the
    nominal parameter values $S_0$.} \label{fig:opt_hybrid_response}
\end{figure}

The optimization algorithm described in Section~\ref{sec:high}
was applied to find the smallest changes in the active and reactive load values,
active and reactive load voltage exponents, and AVR gains that would result in inability to
recover from the disturbance. Three parameter sets were considered:
\begin{enumerate}
\item[$S_1$:] The set of active and reactive load values, capturing load
uncertainty. (38 parameters)
\item[$S_2$:] The set $S_1$ together with the set of active and reactive
load voltage exponents. This set captures the uncertainty not only in the load levels but also their dynamic characteristics. (76 parameters)
\item [$S_3$:] The union of $S_2$ and the set of AVR gains. This illustrates the influence of controller response on system recoverability. (86 parameters)
\end{enumerate}
%Note that $S_1$ consists of 38 parameters,
%$S_2$ of 76 parameters, and
%$S_3$ of 86 parameters.
The optimization was applied to each of these sets of parameter
values, with the solution tolerance set to $\epsilon = 10^{-5}$.
%chosen for the constraint $G(p) = \epsilon$.
The algorithm converged for all of the parameter sets.
Fig.~\ref{fig:opt_hybrid_response} shows the dynamic
response of the generator~5 relative angle for each of the optimal parameter sets,
as well as for the nominal parameter values (denoted by $S_0$).
Notice that the angle trajectories for the optimal solutions all display similar behavior, in particular a larger initial fluctuation than is the case for the nominal
parameter values. Such behavior is indicative of proximity to instability.

The optimization algorithm determines the parameter values which minimize the (square of the) Euclidean distance from the nominal parameter values $p_0$ to the recovery boundary, which is described by $G(p)=0$. Hence, the optimal value of the objective function measures the safe operating margin of the system. Table~\ref{tab:vals2} shows this optimal value, i.e.,~the safety margin, for each set of parameters. Introducing additional parameters increases the degrees of freedom for the optimal solution, ensuring the optimal value cannot increase. Consequently, the optimal value (safety margin) decreased when the parameter set was expanded from $S_1$ to $S_2$, and again from $S_2$ to $S_3$.

For all the various combinations of parameters, the optimization algorithm converged rapidly (in less than 40 iterations) to a set of parameter values on the recovery boundary which appears to be closest to $p_0$. This is an impressive outcome, given the parameter space dimensions are $38$, $76$ and~$86$.

\begin{table}
  \centering
  \caption{Parameter space dimension, safety margin and iteration count for the optimization algorithm of Section~\ref{sec:high}.} \label{tab:vals2}
\begin{tabular}{|c|c|c|c|}
  \hline
  Parameter Set & Dimension & Safety Margin & Iterations
  \\ \hline
  $S_1$ & $38$ & $0.0107$ & 14 \\ \hline
  $S_2$ & $76$ & $0.0079$ & 31 \\ \hline
  $S_3$ & $86$ & $0.0071$ & 39 \\\hline
\end{tabular}
\end{table}

Overall, the algorithms were successfully applied to the IEEE
39-bus benchmark power system to find a point on the recovery boundary in
one-dimensional parameter space, to numerically trace the recovery
boundary in two-dimensional parameter space, and to find the closest point
on the recovery boundary in higher dimensional parameter space.

\section{Conclusion}\label{sec:conc}

The paper introduces the concept of a recovery boundary as the boundary
in parameter space which separates parameter values that induce
post-disturbance dynamic recovery from those that result in failure to
recover. This then motivates the definition of a safety margin for a
given set of parameter values as the shortest distance (in parameter
space) from the given values to the recovery boundary.

Efficient numerical algorithms with convergence guarantees are
presented for numerically tracing the recovery boundary in
two-dimensional parameter space, and for finding the closest point on
the recovery boundary in high dimensional parameter space. This
closest point establishes the safety margin. Unlike prior
developments, these algorithms are not conservative or approximate,
can include parameters which influence post-disturbance dynamics, do not
require prior knowledge of the controlling unstable equilibrium point,
and can include nonsmooth limits such as controller saturation. They
are not limited to low dimensional parameter space and are well suited
for efficiently computing safety margins of realistic power systems.

The algorithms are supported by a theoretical framework which
established an equivalence between a point on the recovery
  boundary in parameter space and the corresponding post-disturbance initial
  condition on the boundary of the region of attraction in state
  space.  Furthermore, this framework showed that the latter could be
found by varying parameter values so as to minimize the inverse
trajectory sensitivities.  These insights underpin the algorithms
presented in the paper, and provide convergence guarantees for those
algorithms under mild assumptions.

The algorithms were demonstrated on the IEEE 39-bus benchmark
power system, where they were used to numerically trace the recovery boundary
in two-dimensional parameter space and to find safety margins in
high dimensional parameter space. The results revealed an unexpected influence of load voltage dependence on system recovery, insights that would not have been identified
otherwise. The algorithms accurately and efficiently
computed the safety margins in each of 38, 76 and~86 dimensional parameter
spaces.

%Future work will consider the recovery boundaries and safety margins for
%multiple faults simultaneously, and will aim to identify the most influential
%parameters and most dangerous faults in a power system.

\IEEEtriggeratref{9}
\bibliographystyle{IEEEtran}
\bibliography{tpwrs_ref}

\begin{IEEEbiography}
  [{\includegraphics[width=1in,height=1.25in,clip,keepaspectratio]
      {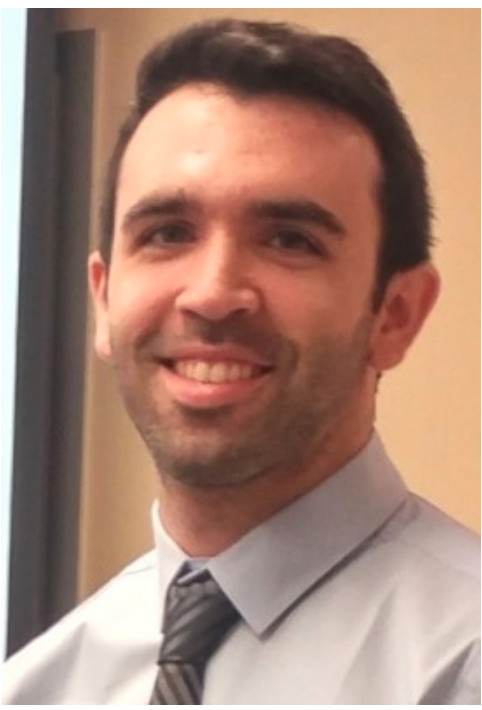}}]{Michael W. Fisher} is an Assistant Professor
  in the Department of Electrical and Computer Engineering at the University of
  Waterloo, Canada.  He was a postdoctoral researcher with
  the Automatic Control and Power System Laboratories at
  ETH Zurich.  He received his Ph.D. in Electrical Engineering:
  Systems at the University of Michigan, Ann Arbor in 2020, and a
  M.Sc. in Mathematics from the same institution in 2017. He received
  his B.A. in Mathematics and Physics from Swarthmore College in 2014.
  His research interests are in dynamics, control, and optimization of
  complex systems, with an emphasis on electric power systems.
  He was a finalist for the 2017 Conference on Decision and Control (CDC)
  Best Student Paper Award and a recipient
  of the 2019 CDC Outstanding Student Paper Award.
\end{IEEEbiography}

\begin{IEEEbiography}
  [{\includegraphics[width=1in,height=1.25in,clip,keepaspectratio]
      {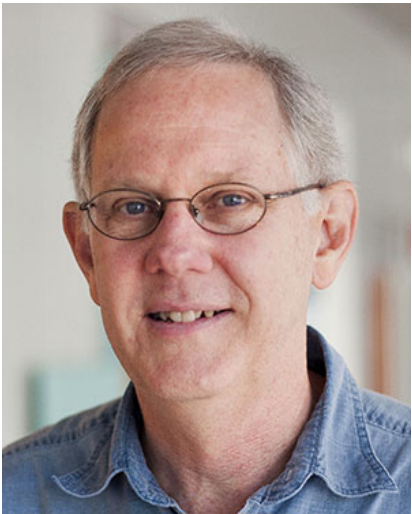}}]{Ian A. Hiskens} is the
  Vennema Professor Emeritus in the Department of Electrical
  Engineering and Computer Science, University of Michigan, Ann Arbor,
  MI, USA. He has held prior appointments with the Queensland Electricity
  Supply Industry, and various universities in Australia and the
  United States. His research interests lie at the intersection of
  power system analysis and systems theory, with recent activity
  focused largely on integration of renewable generation and
  controllable loads.  Dr.~Hiskens is involved in numerous IEEE
  activities in the Power and Energy Society, Control Systems Society,
  Circuits and Systems Society, and Smart Grid Initiative, and was the
  VP-Finance of the IEEE Systems Council. He is the Editor-in-Chief of the
  IEEE Transactions on Power Systems and is a member of the
  Editorial Board of the {\em Proceedings of the IEEE}\@. Dr.~Hiskens is a Life Fellow of IEEE, a
  Fellow of Engineers Australia, a Chartered Professional Engineer in
  Australia, and was the 2020 recipient of the M.A.~Sargent Medal from
  Engineers Australia.
\end{IEEEbiography}

\vfill

% that's all folks
\end{document}